\documentclass[twocolumn,twoside,slac_two]{revtex4}
\usepackage{graphicx}
\usepackage{fancyhdr}
\begin{document}

\title{Development of Multi-Pixel Photon Counters}

%

\author{M.~Yokoyama}\thanks{Corresponding author}\email{masashi@scphys.kyoto-u.ac.jp}
\author{T.~Nobuhara}
\author{M.~Taguchi}
\author{T.~Nakaya}
\affiliation{Department of Physics, Kyoto University, Kyoto 606-8502 Japan}
%
\author{T.~Murakami}
\author{T.~Nakadaira}
\author{K.~Yoshimura}
\affiliation{IPNS, High Energy Accelerator Research Organization (KEK), Tsukuba 305-0801 Japan}
\author{K.~Kawagoe}
\author{Y.~Tamura}
\affiliation{Department of Physics, Kobe University, Kobe 657-8501 Japan}
\author{T.~Iijima}
\author{Y.~Mazuka}
\affiliation{Department of Physics, Nagoya University, Nagoya 464-8601 Japan}
\author{K.~Miyabayashi}
\affiliation{Department of Physics, Nara Women's University, Nara 630-8506 Japan}
\author{S.~Iba}
\author{H.~Miyata}
\affiliation{Department of Physics, Niigata University, Niigata 950-2181 Japan}
\author{T.~Takeshita}
\affiliation{Department of Physics, Shinshu University, Matsumoto 390-8621 Japan}
%

\author{(KEK Detector Technology Project / photon sensor group)}
\noaffiliation

\begin{abstract}
The multi-pixel photon counter (MPPC) is a newly developed photodetector with
an excellent photon counting capability.
It also has many attractive features such as small size, high gain,
low operation voltage and power consumption, and capability of operating in magnetic fields and in room temperature.
The basic performance of samples has been measured.
A gain of $\sim$10$^6$ is achieved with a noise rate less than 1~MHz
with 1~p.e. threshold,  and cross-talk probability of less than 30\% at room temperature.
The photon detection efficiency for green light is twice or more that of the photomultiplier tubes.
It is found that the basic performance of the MPPC is satisfactory for use in real experiments.
\end{abstract}


\maketitle

\thispagestyle{fancy}



\section{MULTI-PIXEL PHOTON COUNTERS}
\begin{figure}[htbp]
\begin{center}
\includegraphics[width=0.23\textwidth]{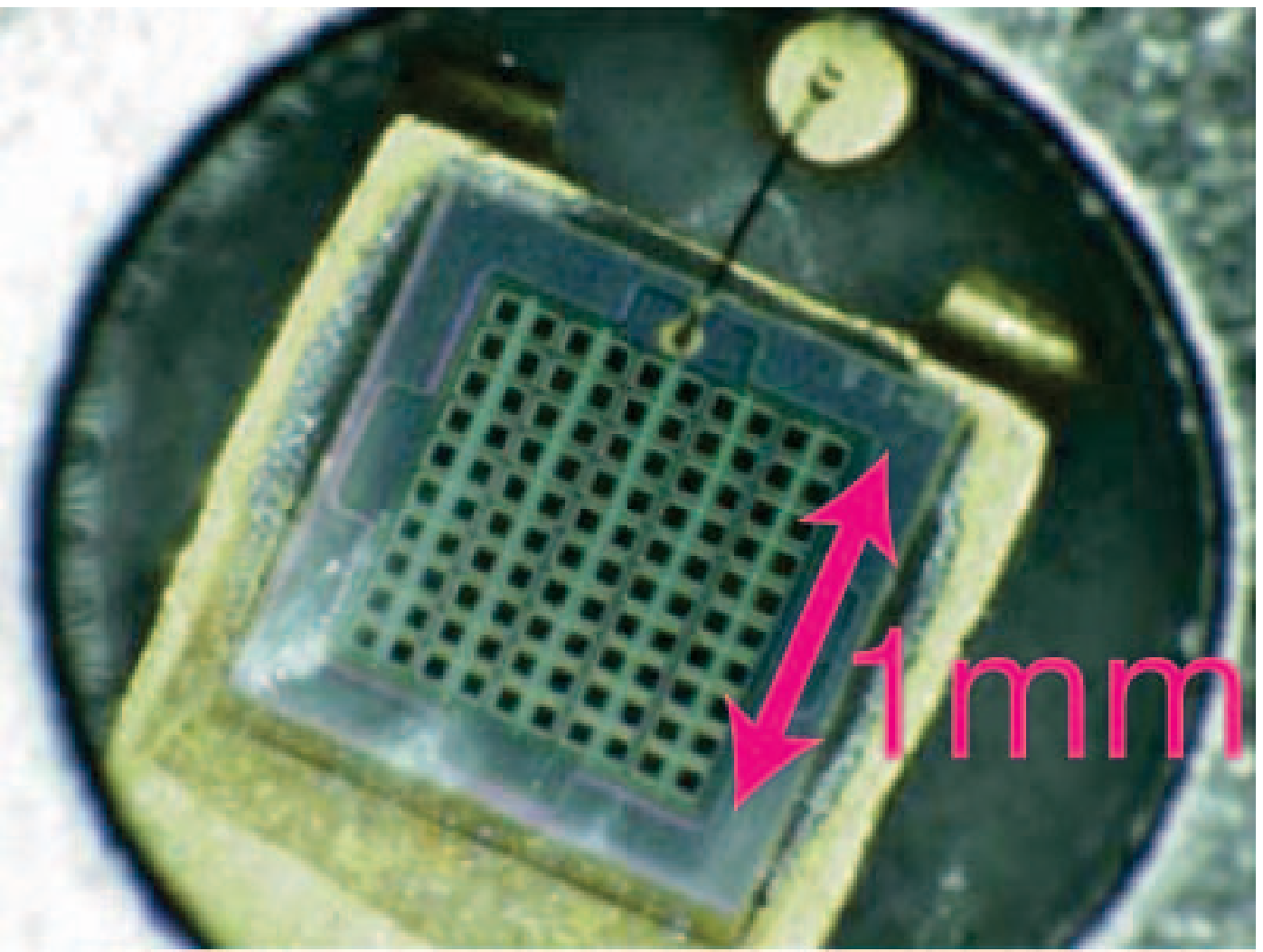}
\includegraphics[width=0.23\textwidth]{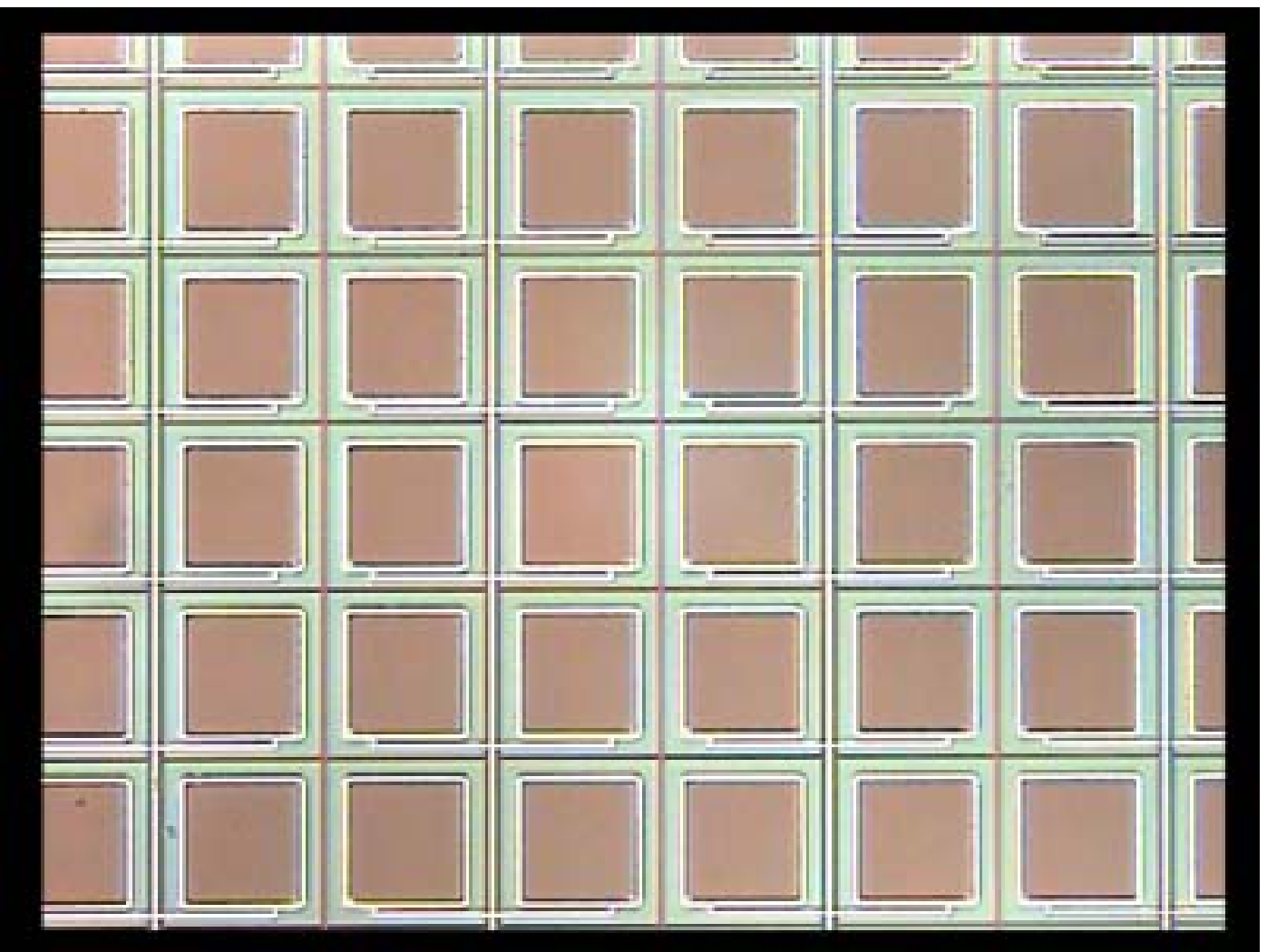}
\caption{
Photograph of a 100-pixel MPPC. Left: entire device in a package. 
Right: close-up view of APD micropixels.
}
\label{fig:MPPC-photo}
\end{center}
\end{figure}

The multi-pixel photon counter (MPPC) is a new photodetector developed
by Hamamatsu Photonics, Japan~\cite{HPK}.  An MPPC consists of many
(100 to $>$1000) small avalanche photodiodes (APDs) in an area of
typically 1~mm$^2$.  Figure~\ref{fig:MPPC-photo} shows a picture of an
MPPC with 100 pixels.

Each APD micropixel independently works in limited Geiger mode
with an applied voltage a few volts above the breakdown voltage ($V_\mathrm{bd}$).
When a photoelectron is produced, it induces a Geiger avalanche.
The avalanche is passively quenched by a resistor integral to each pixel.
The output charge $Q$ from a single pixel
 is independent of the number of produced photoelectrons within the pixel, and
can be written as
$$
Q = C (V-V_\mathrm{bd}),
$$
where $V$ is the applied voltage and $C$ is the capacitance of the pixel.
Combining the output from all the pixels, the total charge from an MPPC is
quantized to multiples of $Q$ and 
proportional to the number of pixels that underwent Geiger discharge (``\textit{fired}'').
The number of fired pixels is proportional
to the number of injected photons if the number of photons is small compared to the total number of pixels.
Thus, the MPPC has an excellent photon counting capability.

For the MPPC, the operation voltage $V$ is a few volts above the breakdown voltage
and well below 100~V.
The pixel capacitance $C$ is on the order of 10--100~fF, giving a gain of 10$^5$--10$^6$.
These features enable us to read out the signal from the MPPC with simple electronics.
In addition, because the thickness of the amplification region is a few $\mu$m, 
it is insensitive to an applied magnetic field and the response is fast.

Because of these great advantages, several particle physics experiments are going to use the MPPC.
The T2K (Tokai-to-Kamioka) long baseline neutrino oscillation experiment~\cite{T2K},
which is now under construction and will start in 2009, 
has chosen the MPPC as one of the baseline photosensors
for the near neutrino detector complex.
Calorimeters for the International Linear Collider are 
another project that is considering use of the MPPC~\cite{ILC-CAL}.
In these experiments, MPPCs will be used to detect photons from plastic scintillators
guided by wavelength-shifting fibers.
With larger area devices or a light collection system, the MPPC may be used for the aerogel-RICH
particle identification system in a B-factory upgrade~\cite{Super-Belle}. 
The features of MPPCs are suitable not only for particle physics experiments but for 
much wider applications such as astrophysics, space science, material science, or medical instruments.

The MPPC is a newly developed device and its performance is rapidly improving.
In this report, the latest status of the development and future prospects are presented.

\section{TEST SAMPLE}
We have tested two types of the latest samples of MPPCs produced by Hamamatsu Photonics.
The difference between them is the number of pixels---one has 100
pixels with a 100~$\mu$m pitch and the other has 400 pixels with a 50~$\mu$m pitch.
A device with a 25~$\mu$m pitch and 1600 pixels is also being tested, although
not reported here.
The main characteristics of the tested samples are summarized in Table~\ref{tab:samples}.
The time response of the MPPC depends on the quenching resistor 
and capacitance of each pixel.
The numbers shown in Table~\ref{tab:samples} are the typical values.

\begin{table}[htdp]
\caption{Summary of tested samples.}
\begin{center}
\begin{tabular}{ccc} \hline \hline
Number of pixels & 100 & 400 \\ \hline
Pixel pitch ($\mu$m) & 100 & 50 \\ \hline
Area & \multicolumn{2}{c} {1.0$\times$1.0 mm$^2$} \\ \hline
Operation voltage & \multicolumn{2}{c} {69--70~V} \\ \hline
Signal temporal width (ns, typ.) & 40 & 10 \\ \hline
 \hline
\end{tabular}
\end{center}
\label{tab:samples}
\end{table}%

\section{BASIC PERFORMANCE}
The basic performance of the MPPC is measured with light from an LED.
All results shown in this section are for a sample 400 pixel device.

\subsection{Raw signal}
\begin{figure}[htbp]
\begin{center}
\includegraphics[width=0.21\textwidth]{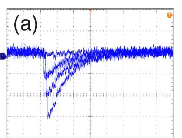}
\includegraphics[width=0.24\textwidth]{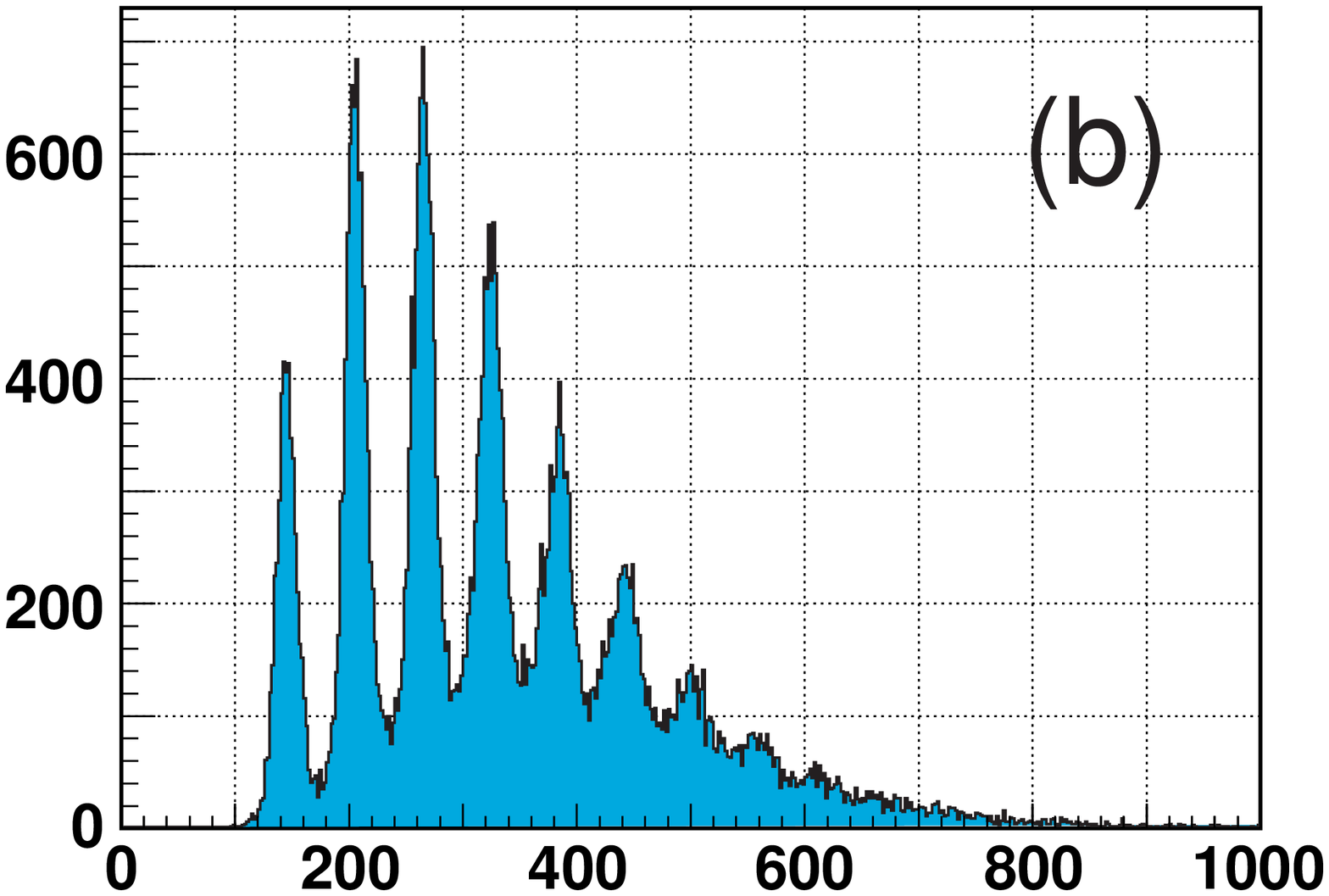}
\caption{
Signal of the MPPC taken with (a) oscilloscope and (b) ADC.
}
\label{fig:raw-signal}
\end{center}
\end{figure}

Figure~\ref{fig:raw-signal}(a) shows the raw signal from an MPPC taken with an oscilloscope.
The MPPC is illuminated by pulsed light from an LED at low intensity and
the oscilloscope is triggered in synch with the LED pulse.
The responses for multiple triggers are overlaid in the figure.
One can see excellently separated signals corresponding to 
one, two, and three fired pixels.
Figure~\ref{fig:raw-signal}(b) shows the output charge read out by an analog-to-digital converter (ADC).
The charge corresponding to different numbers of fired pixels is again well separated (or quantized)
as peaks in the ADC distribution at equal intervals.
This indicates that the gain of each micropixel is uniform in an MPPC.
%
These observations demonstrate the excellent photon counting capability of the MPPC.

\subsection{Gain and noise rate}
\begin{figure}[htbp]
\begin{center}
\includegraphics[width=0.35\textwidth]{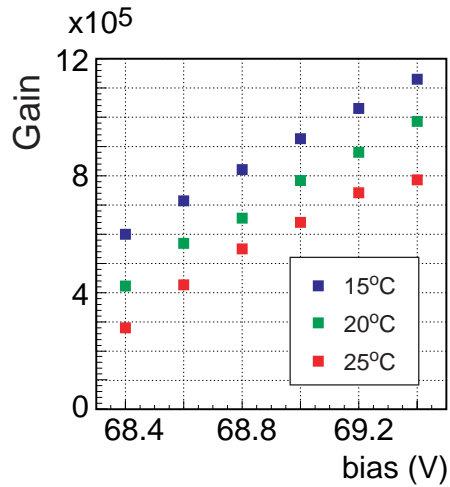}
\caption{
Measured gain as a function of the applied voltage.
Blue, green, and red points correspond to data at 15, 20, and  25$^\circ$C, respectively.
}
\label{fig:gain}
\end{center}
\end{figure}

The gain is measured by illuminating an MPPC with light from an LED.
From the number of ADC counts between
a well-separated pedestal and the peak corresponding to a single fired pixel
(Fig.~\ref{fig:raw-signal}(b)),
we derive the charge corresponding to a single fired pixel, $Q$.
The gain is defined as $Q$ divided by the charge of an electron.
Figure~\ref{fig:gain} shows the measured gain as a function of the applied voltage.
The measurement is performed inside a temperature-controlled chamber and
the data at 15, 20, and  25$^\circ$C are shown.
The measured gain is 3$\times$10$^5$--1.2$\times$10$^6$ and
 linearly depends on the applied voltage as expected.
The breakdown voltage decreases with lower temperature, 
resulting in larger gain at a fixed applied voltage.
The temperature coefficient is about $-$3\%/$^\circ$C.

\begin{figure}[htbp]
\begin{center}
\includegraphics[width=0.35\textwidth]{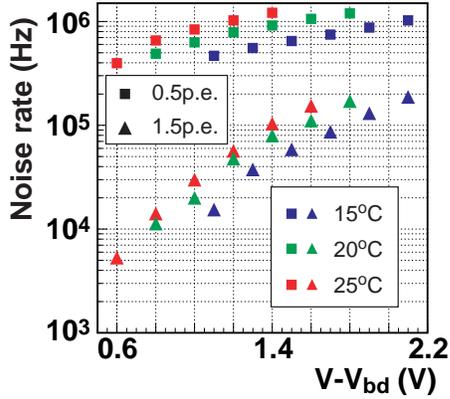}
\caption{
Measured noise rate as a function of $V-V_\mathrm{bd}$.
Blue, green, and red points correspond to data at 15, 20, and  25$^\circ$C, respectively.
Rectangular and triangle points represent the rate with thresholds of 0.5 and 1.5 p.e., respectively.
}
\label{fig:noise}
\end{center}
\end{figure}

The noise rate is measured by counting the rate above a threshold without external light input. 
The results at 15, 20, and  25$^\circ$C are shown in Fig.~\ref{fig:noise}.
The horizontal axis in Fig.~\ref{fig:noise} shows the difference between 
the applied voltage and the breakdown voltage $V_\mathrm{bd}$.
The breakdown voltage is derived by linearly extrapolating the gain-voltage curve in Fig.~\ref{fig:gain} 
to the point where the gain becomes zero.
With a threshold equivalent to 0.5 photoelectrons
(p.e.)\footnote{Here, ``p.e.'' means the number of detected
  photoelectrons, and is assumed to be the same as the number of fired pixels when those numbers are small.}, the noise rate is 0.5--1~MHz.
However, it decreases by about or more than an order of magnitude if the threshold is set to 1.5~p.e.
The noise rate decreases as the temperature becomes lower.
The temperature coefficient of noise rate at 0.5 p.e. threshold is $-7$\%/$^\circ$C.
These observations imply that the dominant component of the noise is due to the discharge of single pixels induced by thermally generated carriers.

We have tested three samples of MPPCs of the same type.
Although the number of the tested devices is limited,
we find that the device-by-device variation of the characteristics is small:
with an applied voltage of 68.8~V, the maximum difference in the gain is less than 3\%,
while that of the noise rate is about 7\%.
We plan to test much larger numbers of devices in the near future.

\subsection{Cross-talk probability}

\begin{figure}[htbp]
\begin{center}
\includegraphics[width=0.35\textwidth]{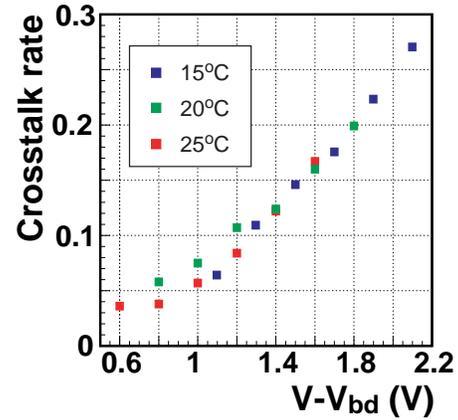}
\caption{
Cross-talk rate as a function of $V-V_\mathrm{bd}$.
Blue, green, and red points correspond to data with 15, 20, and  25~$^o$C, respectively.
}
\label{fig:crosstalk}
\end{center}
\end{figure}

The cross-talk between neighboring micropixels has been observed. 
The origin of the cross-talk is presumed to be optical photons emitted during avalanche~\cite{crosstalk-ref}
which enter neighboring micropixels and trigger another Geiger discharge.

The probability of causing cross-talk is estimated from the fraction of events with more than one p.e. to 
that with one p.e. in randomly triggered events without external light.
We assume that the events with more than one p.e. are caused by the cross-talk from the original Geiger discharge in a single pixel.
The effect from accidental coincidence of two independent discharges is estimated from the fraction of
pedestal events, assuming a Poisson distribution for the original
Geiger discharge, and has been subtracted.

Figure~\ref{fig:crosstalk} shows the cross-talk probability as a function of 
the applied voltage above $V_\mathrm{bd}$.
The cross-talk probability is found to be almost independent of the temperature.
However, it depends on the applied voltage.
The measured cross-talk probability is at an acceptable level for
certain applications (e.g. for the T2K neutrino experiment).
However, for applications that require good linearity with wide dynamic range (e.g. ILC calorimeters), 
this may limit the performance of the device, and improvement may be necessary.

\subsection{Photon detection efficiency}
The photon detection efficiency (PDE) is an important parameter for the performance of the MPPC.
For an MPPC, the PDE is written as a product of three effects:
$$
\mathrm{PDE} = \varepsilon_\mathrm{geom} \times \mathrm{QE} \times \varepsilon_\mathrm{Geiger}.
$$
The geometrical efficiency $\varepsilon_\mathrm{geom}$ represents the fraction of active area in
a micropixel.
It depends on the design and the size of a pixel, but is about 0.5 for current samples. 
The quantum efficiency of the APD, QE, depends on the wavelength of
photon and is typically 0.7--0.8 in the range of current interest.
The probability of inducing a Geiger discharge when a photoelectron is generated, 
$\varepsilon_\mathrm{Geiger}$, depends on the applied voltage.

\begin{figure}[htbp]
\begin{center}
\includegraphics[width=0.45\textwidth]{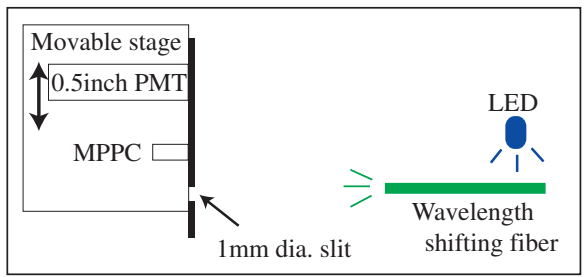}
\caption{
Setup for the PDE measurement.
}
\label{fig:PDE-setup}
\end{center}
\end{figure}

\begin{figure}[htbp]
\begin{center}
\includegraphics[width=0.35\textwidth]{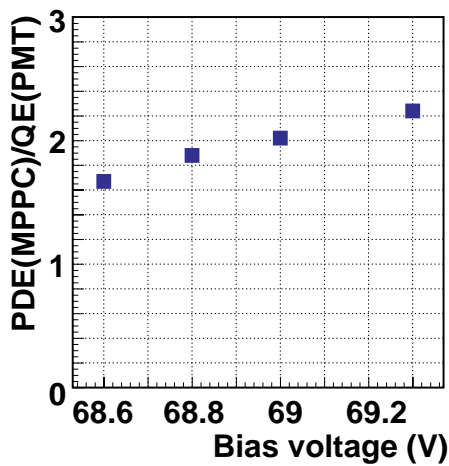}
\caption{
PDE of an MPPC relative to the QE of a PMT as a function of the applied voltage.
}
\label{fig:PDE}
\end{center}
\end{figure}

We have measured the PDE of an MPPC relative to that of a PMT 
using a setup shown in Fig.~\ref{fig:PDE-setup}.
An MPPC and a PMT are illuminated with
green light from a wavelength shifting fiber, Kuraray Y11,
through a slit with 1~mm diameter.
We use Hamamatsu PMT H3165-10 with bialkali photo-cathode as a reference.
Comparing the number of detected photo-electrons between 
the MPPC and the PMT, the relative PDE is measured.
In order to avoid the effect of cross-talk, the number of photoelectrons for the MPPC is derived 
from the fraction of pedestal (= 0~p.e.) events to the total number of trigger, assuming 
a Poisson distribution.
For the PMT, the number of photoelectrons is calculated by dividing the mean output charge by 
the charge corresponding to 1.~p.e.

Figure~\ref{fig:PDE} shows the results of the measurement at 15$^\circ$C.
The PDE of the MPPC for green light is about or more than twice the QE of the PMT.
A measurement of the absolute PDE with a calibrated photodiode will be done in the near future.
 
\subsection{Response to large numbers of photons}

\begin{figure}[htbp]
\begin{center}
\includegraphics[width=0.35\textwidth]{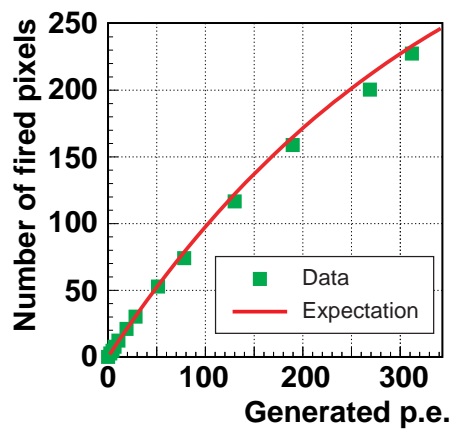}
\caption{
Response of an MPPC as a function of light intensity.
}
\label{fig:linearity}
\end{center}
\end{figure}

The linearity of the MPPC response to a large amount of light is intrinsically limited by 
the finite number of micropixels.
Figure~\ref{fig:linearity} shows the number of fired pixels of an MPPC
as a function of light intensity.
The number of generated photoelectrons inside the active area is 
estimated by  
the intensity of the light monitored by a PMT.
The red curve shows the expected response of the MPPC 
based on the known number of
pixels and the measured cross-talk probability.
The deviation from the expected curve is found to be within 5\%.
Thus, the response of the MPPC is well described by the number of pixels and 
measured cross-talk rate.

\section{TEST WITH A LASER INJECTION SYSTEM}
In order to check the response of each micro APD pixel,
we have tested the MPPC with a pulsed laser system.
The laser has a spot size on the surface of an MPPC
with a diameter of about 10~$\mu$m
so that the uniformity of the response inside a micropixel can be studied.
The MPPC is placed on a X-Y movable stage that can be controlled with 1~$\mu$m accuracy.
The wavelength and the temporal width of the laser are 825~nm
and 50~ps, respectively.
For this test, a sample MPPC with 100 pixels is used.

\subsection{Uniformity inside a micro APD pixel}

\begin{figure}[htbp]
\begin{center}
\includegraphics[width=0.35\textwidth]{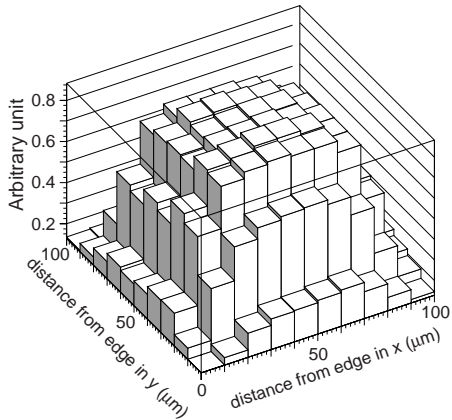}
\caption{
Measured relative response to a light spot across a pixel (100$\times$100~$\mu$m).
}
\label{fig:laser-1pix}
\end{center}
\end{figure}

First, the dependence of the response to the position within a micropixel is studied.
The light spot is injected onto a pixel and 
the area within a pixel is scanned with 10~$\mu$m pitch.
Figure~\ref{fig:laser-1pix} shows the measured response for one pixel.
Because we do not know the number of injected photons in this measurement, 
the $z$ axis is arbitrary and show the relative response.
It is found that the response is uniform within the active area of the micropixel. 

\subsection{Pixel-to-pixel uniformity}

\begin{figure}[htbp]
\begin{center}
\includegraphics[width=0.35\textwidth]{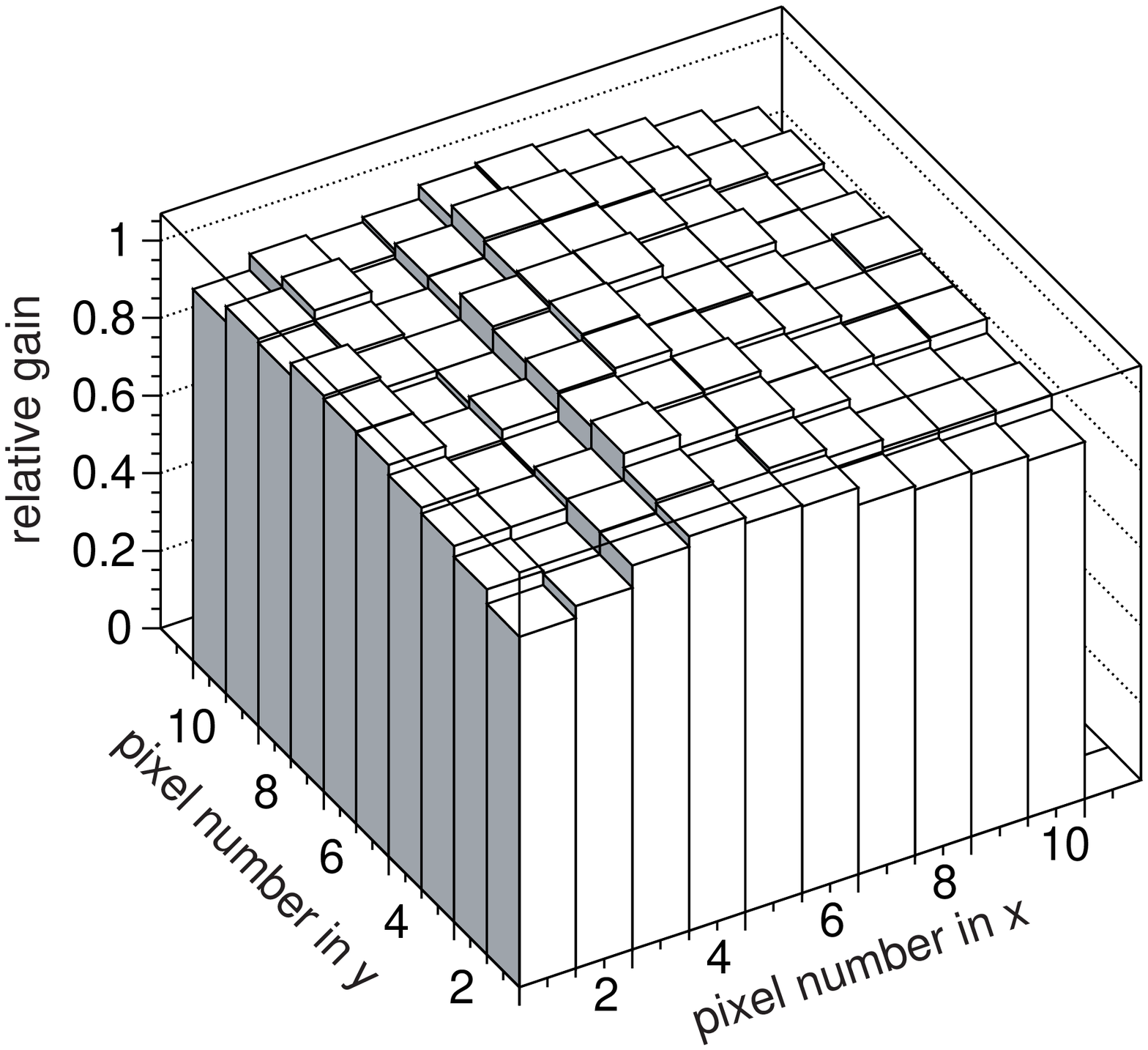}
\includegraphics[width=0.35\textwidth]{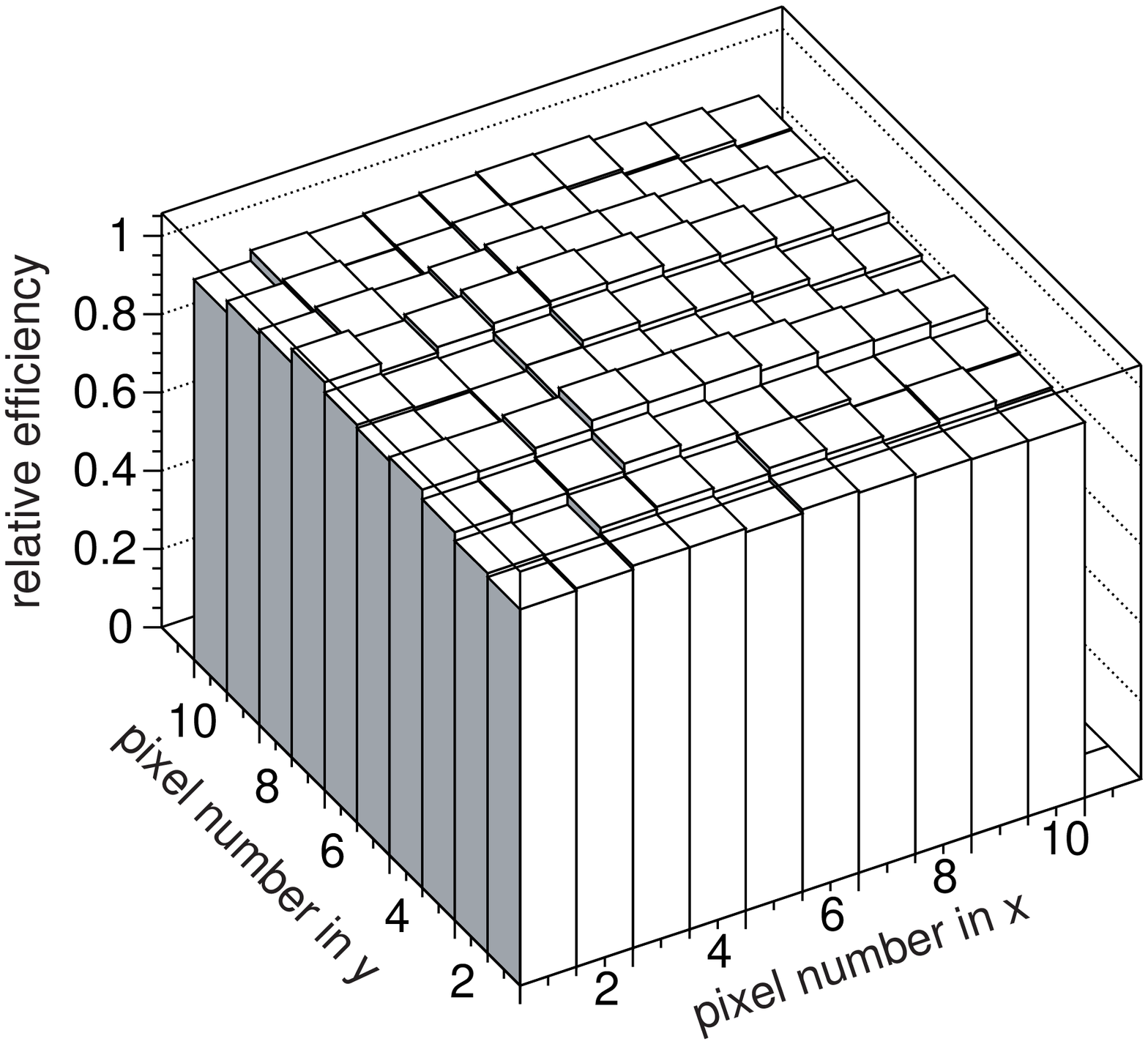}
\caption{
Relative gain (top) and efficiency (bottom) for a 100 pixel device
measured with a laser system.
}
\label{fig:laser-all}
\end{center}
\end{figure}

The pixel-to-pixel uniformity of the gain and the relative efficiency are 
tested by injecting light in the center of each pixel.
The results are shown in Fig.~\ref{fig:laser-all}.
The $z$ axis is normalized so that the average over all pixels is 1.0.
The gain and relative efficiency are found to be uniform within 3.6\% and 2.5\% in RMS, 
respectively.

\section{SUMMARY AND PROSPECTS}
The multi-pixel photon counter (MPPC) is a newly developed photodetector with
an excellent photon counting capability.
It also has many attractive features such as small size, high gain,
low operation voltage and power consumption, and capability of operation in magnetic fields.

We have shown that the basic performance of the MPPC is promising and
actually already satisfactory for use in real experiments:
a gain of $\sim$10$^6$ is achieved with noise rate less than 1~MHz with 1~p.e. threshold and cross-talk probability of less than 30\% at room temperature.
The photon detection efficiency for green light is twice or more that
of a photomultiplier tube.

The development of MPPCs has advanced  well in the past $\sim$2 years,
and the basic functionality of the device has been established.
We will continue the R\&D for more practical development aiming at real use as well as wider application.
The future development will include realization of devices with (but not limited to):
\begin{itemize}
\item larger area,
\item reduced cross-talk,
\item a larger number of pixels, and
\item better photon detection efficiency.
\end{itemize}
Also, the following issues must be investigated for practical use.
\begin{itemize}
\item long term stability, 
\item quality control under mass production,
\item development of suitable readout electronics, and
\item packaging and coupling to radiator.
\end{itemize}
In addition, for wider application we need to investigate:
\begin{itemize}
\item radiation hardness,
\item timing resolution, and
\item high rate capability.
\end{itemize}
Although there remains some work necessary before real application,
we already have plans for most of this work in the near future.

The MPPC is one of the most attractive devices for photon detection
in future high energy physics, astrophysics, material science and medical uses.

\bigskip 
\begin{acknowledgments}
The authors wish to thank the support from the KEK management for the detector technology project.
We are thankful to T.~Yoshida for useful suggestions as a reviewer of the project.
We acknowledge the semiconductor division of Hamamatsu Photonics
K.~K. for providing us with test samples.
We express their special thanks to National Institute of
Informatics for their support of Super-SINET, and to KEK Computing
Research Center for HEPnet-J, which enabled efficient information sharing.
We are grateful to S.~Oser for careful reading of this manuscript and useful discussions.
This work was supported in part by a Grand-in-Aid for
Scientific Research from Ministry of Education,
Science and Technology under Contract No.
17340071.

\end{acknowledgments}

\bigskip 

\end{document}